\documentclass[twocolumn]{aastex62}

\usepackage{graphicx}
\usepackage{xspace}
\usepackage{xcolor}
\usepackage{amsmath}

\newcommand{\secref}[1]{Section \ref{#1}}
\newcommand{\figref}[2]{Figure \ref{#1}{#2}}
\newcommand{\unit}[1]{\,\ensuremath{\mathrm{#1}}}
\renewcommand{\vec}[1]{\ensuremath{\mathbf{#1}}}

\newcommand{\Halpha}{\ensuremath{\mathrm{H_\alpha}}\xspace}

\newcommand{\etc}{\textsl{etc}\xspace}
\newcommand{\ie}{\textsl{i}.\textsl{e}.\xspace}
\newcommand{\eg}{\textsl{e}.\textsl{g}.\xspace}

\newcommand{\ydquote}[1]{``#1"}

\received{ }
\revised{ }
\accepted{ }
\submitjournal{ApJL}

\shorttitle{Kelvin Helmholtz Instability}
\shortauthors{Yuan et al.}
\begin{document}

\title{Multi-layered Kelvin-Helmholtz Instability in the Solar Corona}

\correspondingauthor{Ding Yuan}
\email{yuanding@hit.edu.cn; ydshen@ynao.ac.cn}

\author[0000-0002-9514-6402]{Ding Yuan}
\altaffiliation{These authors contribute equally to this work}
\affil{Institute of Space Science and Applied Technology, \\
	   Harbin Institute of Technology, Shenzhen 518055, China}
\affil{Key Laboratory of Solar Activity, National Astronomical Observatories,\\ 
	   Chinese Academy of Sciences, Beijing 100012, China}

\author{Yuandeng Shen}
\altaffiliation{These authors contribute equally to this work}
\affil{Yunnan Astronomical Observatory, Chinese Academy of Sciences, \\
	         PO Box 110, Kunming 650011, China}
      
\author{Yu Liu}
\affiliation{Yunnan Astronomical Observatory, Chinese Academy of Sciences, \\
	PO Box 110, Kunming 650011, China}
\author{Hongbo Li}
\affil{Institute of Space Physics, Luoyang Normal University, Luoyang 471934, China}

\author{Xueshang Feng}
\affil{Institute of Space Science and Applied Technology, \\
	Harbin Institute of Technology, Shenzhen 518055, China}

\author{Rony Keppens}
\affil{Centre for mathematical Plasma Astrophysics, Department of Mathematics, \\ 
	 KU Leuven, Celestijnenlaan 200B, Leuven 3001, Belgium}
\affil{School of Astronomy and Space Science, Nanjing University, Nanjing 210023, China}
\affil{Purple Mountain Observatory, Chinese Academy of Sciences, Nanjing 210023, China}


\begin{abstract}
The Kelvin-Helmholtz (KH) instability is commonly found in many astrophysical, laboratory, and space plasmas. It could mix plasma components of different properties and convert dynamic fluid energy from large scale structure to smaller ones. In this study, we combined the ground-based New Vacuum Solar Telescope (NVST) and the Solar Dynamic Observatories (SDO) / Atmospheric Imaging Assembly (AIA) to observe the plasma dynamics associated with active region 12673 on 09 September 2017. In this multi-temperature view, we identified three adjacent layers of plasma flowing at different speeds, and detected KH instabilities at their interfaces. We could unambiguously track a typical KH vortex and measure its motion. We found that the speed of this vortex suddenly tripled at a certain stage. This acceleration was synchronized with the enhancements in emission measure and average intensity of the 193 \AA{} data. We interpret this as evidence that KH instability triggers plasma heating. The intriguing feature in this event is that the KH instability observed in the NVST channel was nearly complementary to that in the AIA 193 \AA{}. Such a multi-thermal energy exchange process is easily overlooked in previous studies, as the cold plasma component is usually not visible in the extreme ultraviolet channels that are only sensitive to high temperature plasma emissions. Our finding indicates that embedded cold layers could interact with hot plasma as invisible matters. We speculate that this process could occur at a variety of length scales and could contribute to plasma heating.
\end{abstract}



\section{Introduction} 
\label{sec:intro}
The Kelvin-Helmholtz (KH) instabilities can occur in any fluid or plasma with a continuous velocity shear or at the interface of two shearing fluids with different density or temperature \citep{Kelvin1871,Helmholtz1868}. However, in a magnetized plasma, compressibility and magnetic tension could have a stabilizing effect on the instabilities. Hence, the velocity shear has to reach a threshold to grow into instability \citep{Chandrasekhar1961}. 

The KH instability is an important mechanism in the evolution of turbulence in the stratified interior of the ocean \citep[\eg,][]{Smyth2012} and in the atmosphere of the Earth and giant planets, \eg, Jupiter, Saturn \citep[\eg,][]{Houze2014}. KH instability is also detected in collisionless space plasmas throughout the solar system, for instance, at the magnetopause of the Earth, Mercury, Jupiter and Saturn \citep[\eg,][]{Johnson2014}, and in many space and astrophysical plasmas \citep[\eg,][]{Murray1993,Vietri1997,Wang2001,Lobanov2001,Bucciantini2005,Berne2010}.

In the solar atmosphere, KH instability is observed at a variety of scales, \eg, as growing ripples at the interface between a prominence and the corona \citep{Ryutova2010,Berger2017,Yang2018,Hillier2018}. KH and other plasma instabilities are believed to be the key processes in dispersing and evaporating cool prominence material into the hot corona \citep{Berger2017,Hillier2018,LiDong2018}. \citet{Ofman2011} reported the growth and saturation of KH vortices at the interface between erupting and non-erupting plasmas during a coronal material ejection (CME) event. Similar KH vortices were observed at the flank of an erupting CME \citep{Foullon2011} and coronal streamers \citep{Feng2013}. 

In high temperature plasma, field-aligned conductivity is very large \citep{Braginskii1965}, the charged particles are frozen in the magnetic field lines, \ie, the plasma expands and contracts with conserved magnetic flux. In typical low $\beta$ coronal plasma, strong magnetization ensures that plasma could stream freely along the magnetic field lines, so we expect any velocity gradient would form preferentially across the magnetic field. \citet{LiXiaohong2018} reported plasma temperature  enhancement after KH instability in coronal loops. It implies that KH instability could trigger plasma heating. This process was elucidated by nonlinear magnetohydrodynamic (MHD) simulations \citep[\eg,][]{Fang2016}. Recently, \citet{Ruan2018} simulated the growth of KH instability in post-flare loop, which was invaded by evaporation flows. Loop-top soft and hard X-ray emission sources was predicted. KH vortices thereby grow into a highly nonlinear stage and roll up the magnetic field lines. Magnetic islands are formed and release energy by magnetic reconnection \citep{Fang2016,Ruan2018}.

In this study, we report the observation of multi-thermal layers that interact with one another by means of KH instability. The interaction between cool and hot plasma sheets results in localized temperature enhancements and acceleration of bulk plasma. The observation and method are given in \secref{sec:obs}; the results are presented in \secref{sec:results}; then we proceed to discussions and conclusions in \secref{sec:con}.

\section{Observation and data analysis} 
\label{sec:obs}

\begin{figure*}[ht]
\centering
\includegraphics[width=0.7\textwidth]{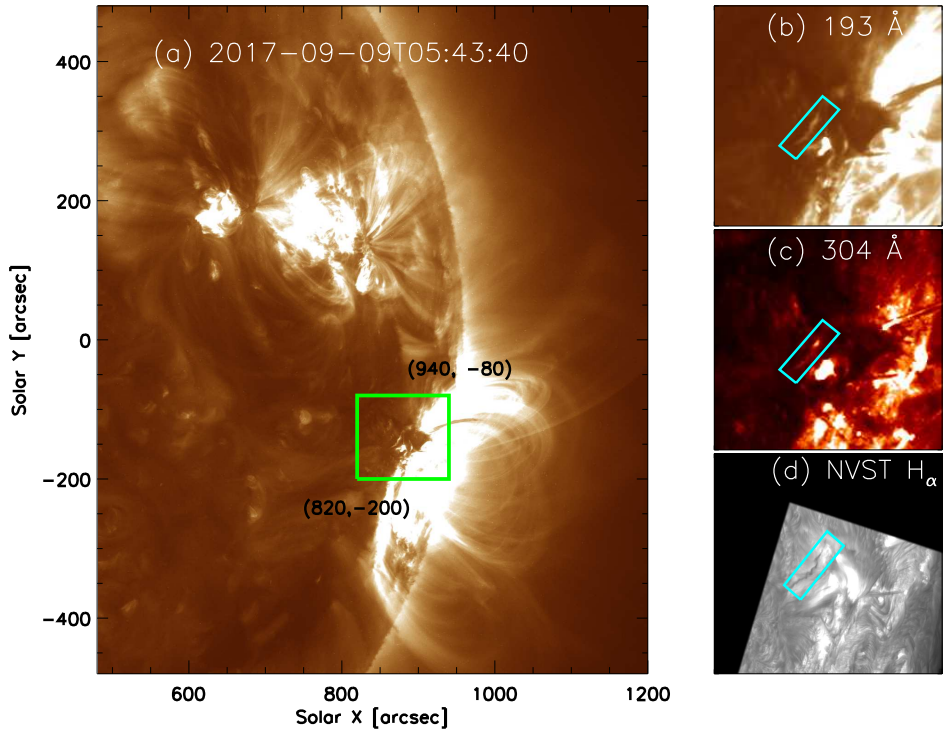}
\caption{(a) Field-of-View of the AIA 193 \AA{} channel showing the plasma motion on the north-east outskirt of AR12673. (b) - (d) highlight the region of interest (the green box in panel a) in the 193 \AA{}, 304 \AA{} and NVST \Halpha images, respectively. The area enclosed by the tilted rectangle are erected in \figref{fig:vortex} to visualize the plasma motion. (An animation of this figure is available.) \label{fig:fov}}
\end{figure*}

A GOES class M1.1 flare was observed at the active region (AR) 12673 on 09 September 2017 by the Atmospheric Imaging Assembly \citep[AIA,][]{Lemen2012} on board the Solar Dynamics Observatories (SDO). This flare was triggered on 04:14:00 UT and stopped on 04:43:00 UT. The New Vacuum Solar Telescope \citep[NVST,][]{Liu2014} operated between 05:34:14 UT and 06:07:30 UT, so it only recorded the relaxation stage of the flare. A bulk of plasma erupted after the flare. In the meantime, magnetic field lines relaxed from intertwinement. A filament-like plasma sheet was left over, presumably being supported by the relaxed magnetic field, which appears to extend radially into outer space (see \figref{fig:fov}{a}). This plasma sheet moved horizontally across the dominant magnetic field, and developed a chain of kink displacements in a snake-like shape (see \figref{fig:fov}{b-c}).

Our observation was made with the ground-based high-resolution NVST and SDO/AIA. In NVST's operation, the H$_\alpha$ Lyot filter was tuned to the line center ($\lambda=6562.8\unit{\AA{}}$) for fast imaging. This narrow-band filter was optimized to record plasma emissions at about 10,000 K. The filter's bandwidth is about $0.25\unit{\AA{}}$, each image was recorded with an exposure time of about 20 ms. The sampling interval was about 5 seconds, and the spatial resolution 0.262 arcsec or about 190 km. SDO/AIA took extreme ultraviolet (EUV) images about every 12 seconds with a spatial resolution of about 1.2 arcsec. AIA EUV filters was optimized to record the emissions of hot plasma with temperature ranging from 50,000 K to 20, 000,000 K.

We processed the NVST \Halpha images by removing the dark current and normalizing them with a flat field. Then, we applied a lucky imaging algorithm to the data. Finally, the NVST images were rotated to align with the solar North and translated to match the key features recorded in the AIA 304 \AA{} channel (see \figref{fig:fov}{c}). The AIA images were calibrated with the standard processing routines available in the solar software library \citep{Freeland1998}.  

We combined multi-wavelength EUV imaging data recorded by SDO/AIA: 94 \AA{}, 131 \AA{}, 171 \AA{}, 193 \AA{}, 211 \AA{} and 335 \AA{}, and used a regularized inversion method \citep{Hannah2012} to recover Differential Emission Measure (DEM). The DEM of a coronal element normally varies with temperature in a Gaussian profile \citep{DelZanna2015}, therefore, the plasma temperature can be estimated as the value where the DEM reaches its maximum. The emission measure (EM) was calculated by integrating DEM over temperature. The EM is proportional to the electron density squared, \ie, $\mathrm{EM}=0.83hn^2_\mathrm{e}$, where $h$ is the column depth, $n_\mathrm{e}$ is the number density of electron, and the factor of 0.83 arises from accounting contribution of the ionized helium electrons.  

\begin{figure*}[ht]
\centering
\includegraphics[width=0.7\textwidth]{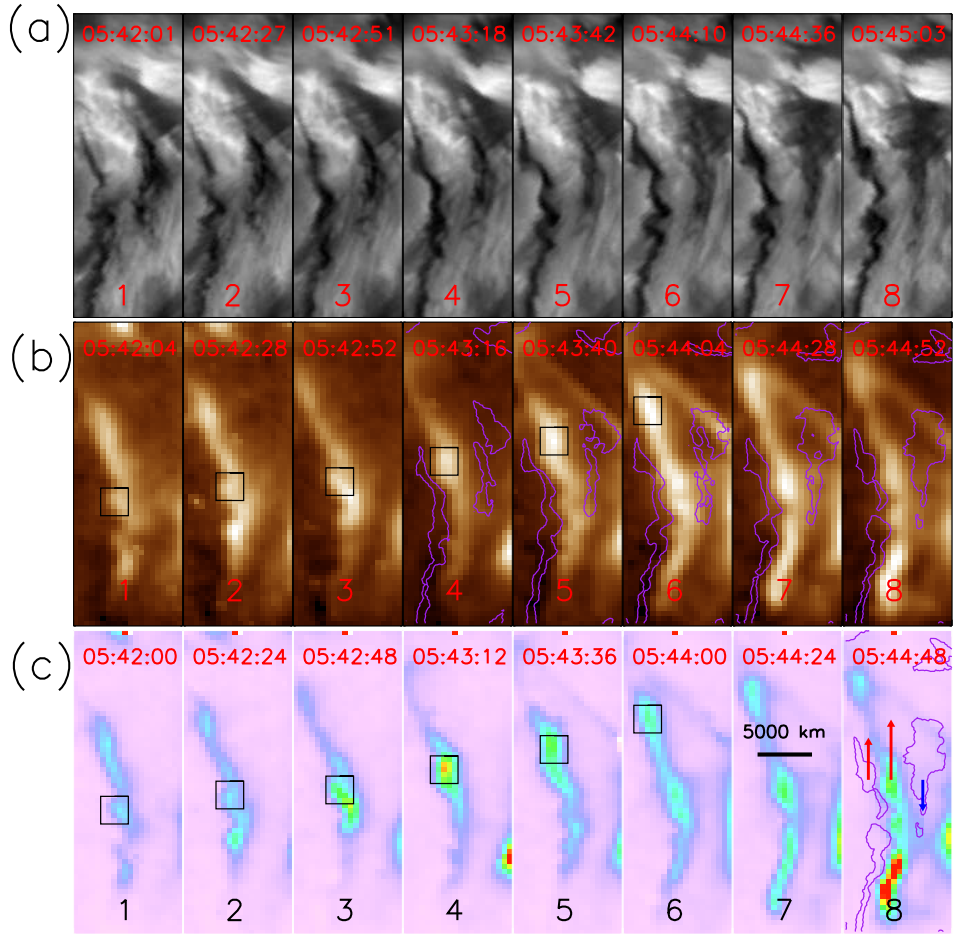} 
\caption{(a) The plasma flow evolution observed by the NVST \Halpha channel. (b) The coronal flow component observed by the AIA 193 \AA{} channel. The contours overlaid in 4th-8th snapshots are based on the NVST H$_\alpha$ emission intensities displayed in panel (a). It shows that boundaries of the coronal and chromospheric flow component are complement each other at the contact surface. (c) The DEM at T=1,250,000 K. Uniform time stamps of 1-8 are allocated to each snapshot, although the measurements for each instrument are taken at slightly shifted times. The contour and arrows plotted in the $8{\mathrm{th}}$ frame in panel (c) gives the relative motions of the plasma material, the lengths of arrows are scaled with the speed value, \ie, $28\unit{km\cdot s^{-1}}$, $40\unit{km\cdot s^{-1}}$ and $-21\unit{km\cdot s^{-1}}$, respectively. \label{fig:vortex}}
\end{figure*}

\begin{figure}[ht]
	\centering
	\includegraphics[width=0.5\textwidth]{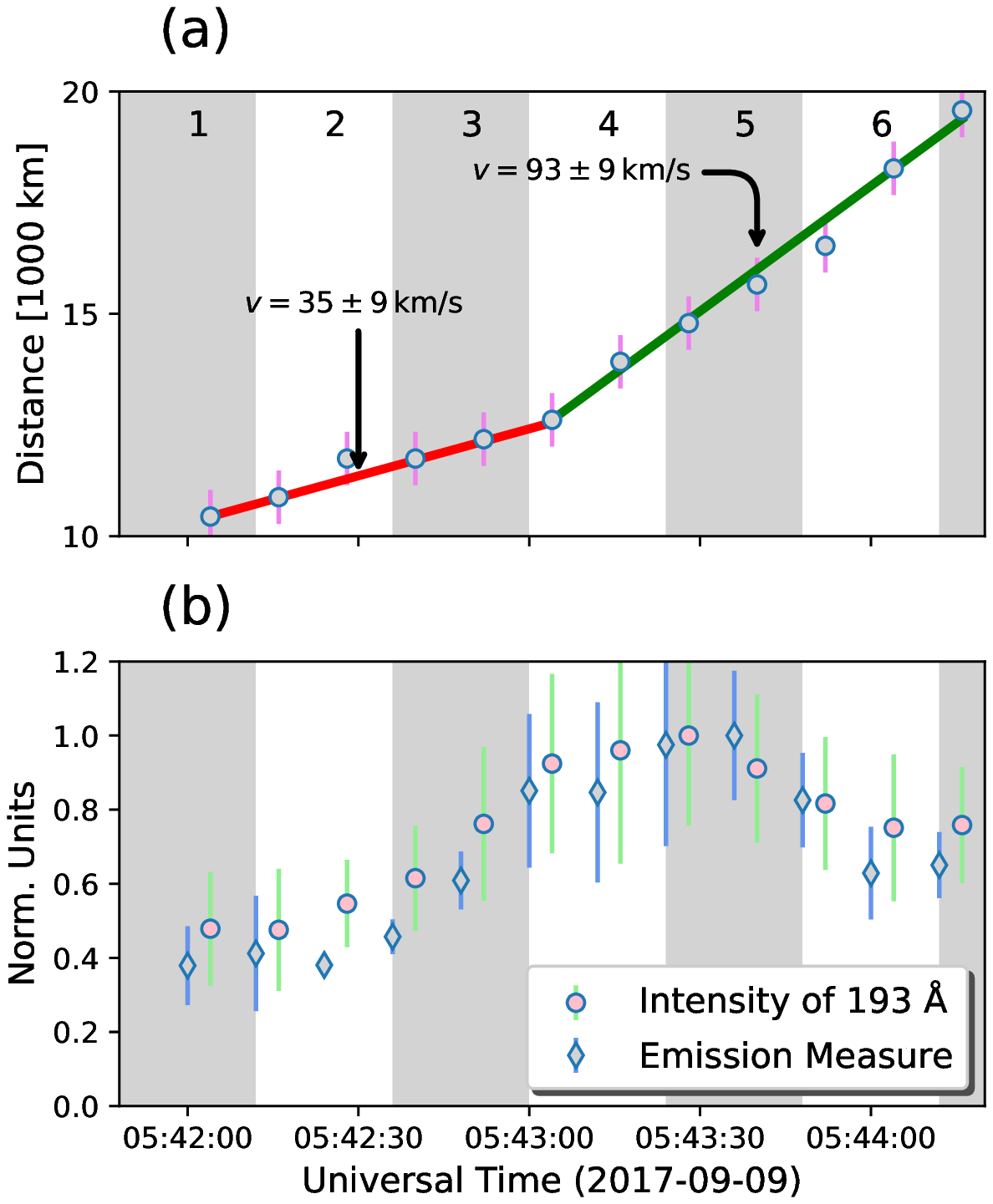} % this command will be ignored
	\caption{(a) The position of the plasma blob traced in \figref{fig:vortex}. (b) The average emission intensity in the AIA 193 \AA{} channel and the EM. The time stamps correspond to those labeled in \figref{fig:vortex}. \label{fig:heating}}
\end{figure}

In order to study the evolution of the instability, we traced a blob of plasma, its positions are marked in \figref{fig:vortex}{c}. We used the pixel with maximum emission intensity within each box as the barycenter, the tracking error was estimated to be one AIA pixel. We spotted the difference in propagation speed before and after 05:43:12 UT (time stamp 4), so we used two linear fits to obtain the speeds, the result is illustrated in \figref{fig:heating}{a}.

\section{Results}
\label{sec:results}

\subsection{Parameters for the KH instability}
The KH instability initially developed as a kink displacement of the plasma filament, as observed in most AIA channels and in detail by the NVST H$_\alpha$ channel. The difference between a stable wave and a linear KH instability relies on whether the growth rate of wave amplitude is zero or positive. In our case, the wave amplitude grew. The wavelength was about $2\unit{Mm}$. NVST revealed the fine detail of the kink motion: the wavelength varied from $1\unit{Mm}$ to $2.5\unit{Mm}$ along the spine of the filament (\figref{fig:vortex}). The vortices also varied in size from less than $1\unit{Mm}$ to about $2\unit{Mm}$. The propagation speed was about $30\unit{km\cdot s^{-1}}$. 
% So the period of KH instability mode is about $30-80\unit{s}$, which situates in the short-period part of fast kink MHD mode statistics \citep{Goddard2016}. 
% If we consider that the plasma sheet is consist of chromospheric material with higher density, and that local \Alfven speed is smaller than that of ambient corona, this period is a reasonable value.

\subsection{Multi-layered KH instability}

Two flow components were observed in the NVST H$_\alpha$ channel (\figref{fig:vortex}a). They propagated towards opposite directions, the projected propagation speeds were about 28 $\mathrm{km\,s^{-1}}$ and -21 $\mathrm{km\,s^{-1}}$, respectively. These two flows were separated by a layer of invisible material (\figref{fig:vortex}a). We plot the contour of H$_\alpha$ emission over the 193 \AA{} images in \figref{fig:vortex}{b}. It reveals that the flow component in the 193 \AA{} channel fills up the gap, the projected propagation speed of this layer was about 40 $\mathrm{km\,s^{-1}}$. A remarkable feature is that kink displacements associated with the KH instability in \Halpha channel were complementary to the counterpart in the 193\AA{} channel, see \figref{fig:vortex}{b}. It indicates that these three flows captured at different channels were adjacent layers of shearing plasma. This scenario is depicted in the $8^\mathrm{th}$ panel of \figref{fig:vortex}{c}.

\subsection{Onset and growth of KH instability}

AR 12673 had rotated to the limb, so we could not obtain the coronal magnetic field by extrapolation. The onset condition of KH instability at the contact surface of two bulk plasmas is \citep{Chandrasekhar1961}
\begin{equation}
\left[\vec{k}\cdot\Delta\vec{v}\right]^2>\frac{\rho_1+\rho_2}{\mu_0\rho_1\rho_2}
\left[(\vec{k}\cdot\vec{B_1})^2+(\vec{k}\cdot\vec{B_2})^2\right],
\end{equation}
where $\vec{k}=2\pi/\lambda$ is the wave vector, $\lambda$ is the wavelength, $\Delta\vec{v}$ is the velocity difference, $\rho_1$ and $\rho_2$ are the respective density of two layers; $\vec{B_1}$ and $\vec{B_2}$ are the magnetic field vectors in two plasma layers, $\mu_0$ is the magnetic permeability in free space. For simplicity reason, we assume that the magnetic field is the same in two plasmas, \ie, $\vec{B_2}=\vec{B_1}$. The onset condition gives an up limit for the parallel magnetic component, 
\begin{equation}
B_\parallel<\frac{\Delta v \sqrt{\mu_0\rho_2\rho_1}}{\sqrt{2(\rho_2+\rho_1)}}=0.8\unit{G}.
\end{equation}
Here we have used a density ratio $\rho_2/\rho_1=10$ in this estimation. The $B_\parallel$ obtained here is only a fraction of typical coronal magnetic field strength. However we shall note there are a number of factors that we have not considered: (1) \citet{Chandrasekhar1961} assumes the KH instability grows from a small-amplitude linear perturbation to a sharp contact interface and uses incompressible conditions. In real observation, the deformed vortices could well violate first-order perturbation. We may have measured the nonlinear stage of KH instability; (2) A real plasma involves extra physical terms other than ideal MHD, \eg, viscosity, thermal conduction, partial ionization, heating \etc. 

\subsection{Heating effect}
\label{sec:heating}

We traced a vortex structure within a $5\times5$ macro pixel as labeled in \figref{fig:vortex}b, and measured its position, averaged emission intensity and EM. The EM was measured to be $3.7^{\pm1.8}\cdot10^{27}\,\mathrm{cm^{-5}}$, and the temperature $1.5^{\pm0.5}\cdot10^6\unit{K}$. If we assume that the column depth was about 3-5 pixels, namely $h=1.8^{\pm0.6}\cdot10^6\unit{m}$, then the number density of electrons was estimated with $n_\mathrm{e}=\sqrt{\mathrm{EM}/0.83h}=5.0^{\pm1.5}\cdot{10^{10}}\unit{cm^{-3}}$ \citep[also see][]{Aschwanden2013}. This blob of material had a plasma density at the level of flaring loops \citep{Huang2018}. 

The center of this vortex migrated at a speed of about $35^{\pm9}\,\mathrm{km\,s^{-1}}$ and suddenly almost tripled to about $93^{\pm9}\,\mathrm{km\,s^{-1}}$ at 05:43:00 UT (\figref{fig:heating}a). In the mean time, the emission intensity and EM reached their maximums (\figref{fig:heating}b). It is evident that the sudden jump in propagation speed was a response to localized plasma heating, which was likely to be triggered by the KH instability.   

If we assume that the energy deposition was constrained within an area $A=\pi D^2/4$, and D was estimated as 3 AIA pixels (about 1,500 km), and that it was released within $\delta t=12$ seconds, the dynamic energy gained by the plasma vortex was estimated as $\delta E= \delta q A\delta t$, where $q=1/2\rho v^3$ is the energy density flux of dynamic energy in a directional flow. In this case, we assume the inflow and outflow had speeds of $35\unit{km\cdot s^{-1}}$ and $93\unit{km\cdot s^{-1}}$, respectively (see \figref{fig:heating}{a}), and the density remained constant over the course. So the dynamic energy gained during this acceleration was about $5.0\cdot10^{17}$ Joule. This is about 2-3 orders of magnitude smaller than a detectable flare, but 2-5 orders of magnitude greater than a nanoflare \citep{Klimchuk2006}. Sudden energy relaxation impulsively heated the local plasma to high temperature, and resulted in a sudden jump of local gas pressure, which then caused quick expansion and acceleration of local plasma.

\section{Discussions and conclusion}
\label{sec:con}

In this study, we observed that multiple adjacent layers with significant density and temperature contrasts flowed with a velocity shear, and found that KH instabilities grew at their contact surfaces. The sudden jerk of a plasma vortex was synchronized with the enhancements for emission measure of the plasma and the emission intensity of the 193 \AA{} channel. It implies that a sudden energy release may have occurred at that time.

The uniqueness of this event is that if one uses a single narrow band channel that is sensitive to hot plasma emission, one would intuitively neglect the interaction with cold plasma, and vice versa. However, we show in this multi-instrument study, cold plasma interacts with coronal plasma as invisible matter, and bolsters mass and energy exchanges.

This sort of event cannot be observed with a single narrow band channel. However, shearing motions are very common in the solar atmosphere and should occur at a vast range of scales. The dark features in the EUV images are intuitively considered as \ydquote{vacuum}, and therefore, are usually neglected. A positive example is that the dark small-scale filament eruption are found as the driver for X-ray jets and revised the jet eruption model \citep{Shen2012,Sterling2015,Shen2017}. \citet{DePontieu2011} similarly raises the importance of heating by type II spicules at the interface of photospheric and coronal material. As the corona is very inhomogeneous and is filled with dark features at a variety of scales, the induced interaction between hot and cold plasmas could play a significant role in energy dissipation throughout the solar corona.

KH instability and its associated secondary effects could be well observed in space plasmas. During a coronal mass ejection, KH instability could be measured at very limited viewing angles as demonstrated by three-dimensional simulation and forward modelling \citep{Syntelis2019}, so its occurrence could have been under-estimated owing to projection effect. In order to assess the associated heating process and plasma acceleration as observed in this event, one may has to do a three-dimensional simulation with multi-fluid approach. Such kind of events are reported in space plasmas \citep[\eg][]{Moore2017}, a dedicate review on KH instability and its secondary processes in space plasmas could be found in \citet{Masson2018}. In the solar corona, ion-scale processes would not be resolved with current instrumentation in the near future, but the secondary effects of MHD instabilities, such as plasma heating, particle acceleration, and mass and energy transportation, could manifest themselves in the macroscopic plasma parameters during MHD-scale observations.

\acknowledgments
The data used in this paper were obtained with the New Vacuum Solar Telescope in Fuxian Solar Observatory of Yunnan Astronomical Observatory, Chinese Academy of Sciences. DY jointly is supported by the National Natural Science Foundation of China (NSFC, 11803005), Shenzhen Technology Project (JCYJ20180306172239618), and the Open Research Program (KLSA201814) of Key Laboratory of Solar Activity of National Astronomical Observatory of China. YDS is supported by the NSFC (11773068, 11633008, 11922307) and the Yunnan Science Foundation (2017FB006); YL by the NSFC (11533009); RK by FWO-NSFC grant G0E9619N.

\bibliography{yuan2019_khi}



\end{document}